\documentclass[preprint,prb,aps,showpacs,footinbib,amsmath,amssymb,superscriptaddress]{revtex4-1}
\usepackage{graphicx}
\usepackage{dcolumn}
\usepackage{color}
\usepackage{bm}
\usepackage{natbib,hyperref}
\usepackage{amsmath,amssymb,amsfonts,bbold}
\usepackage[normalem]{ulem}

\newcommand{\beq}{\begin{equation}}
\newcommand{\eeq}{\end{equation}}
\newcommand{\beqa}{\begin{eqnarray}}
\newcommand{\eeqa}{\end{eqnarray}}

\graphicspath{{fig/}{theo-fig/}}
 
\begin{document}
\title{On the preparation and electronic properties \\ of clean superconducting Nb(110) surfaces}
\author{Artem~B.~Odobesko}
	\email[corresponding author: \\]{artem.odobesko@physik.uni-wuerzburg.de}
	\affiliation{Physikalisches Institut, Experimentelle Physik II, 
		Universit\"{a}t W\"{u}rzburg, Am Hubland, 97074 W\"{u}rzburg, Germany}
	\affiliation{Kotel'nikov IRE RAS, Mokhovaya 11, 125009 Moscow, Russia}
\author{Soumyajyoti Haldar}
	\affiliation{Institut f\"{u}r Theoretische Physik und Astrophysik, 
		Christian-Albrechts-Universit\"{a}t zu Kiel, Leibnizstr.\,15, 24098 Kiel, Germany}
\author{Stefan Wilfert}
	\affiliation{Physikalisches Institut, Experimentelle Physik II, 
		Universit\"{a}t W\"{u}rzburg, Am Hubland, 97074 W\"{u}rzburg, Germany}
\author{Jakob Hagen}
	\affiliation{Physikalisches Institut, Experimentelle Physik II, 
		Universit\"{a}t W\"{u}rzburg, Am Hubland, 97074 W\"{u}rzburg, Germany}
\author{Johannes Jung}
	\affiliation{Physikalisches Institut, Experimentelle Physik II, 
		Universit\"{a}t W\"{u}rzburg, Am Hubland, 97074 W\"{u}rzburg, Germany}
\author{Niclas Schmidt}
	\affiliation{Physikalisches Institut, Experimentelle Physik II, 
		Universit\"{a}t W\"{u}rzburg, Am Hubland, 97074 W\"{u}rzburg, Germany}
\author{Paolo Sessi}
	\affiliation{Physikalisches Institut, Experimentelle Physik II, 
		Universit\"{a}t W\"{u}rzburg, Am Hubland, 97074 W\"{u}rzburg, Germany}
\author{Matthias Vogt}
	\affiliation{Physikalisches Institut, Experimentelle Physik II, 
		Universit\"{a}t W\"{u}rzburg, Am Hubland, 97074 W\"{u}rzburg, Germany}
\author{Stefan Heinze}
	\affiliation{Institut f\"{u}r Theoretische Physik und Astrophysik, 
		Christian-Albrechts-Universit\"{a}t zu Kiel, Leibnizstr.\,15, 24098 Kiel, Germany}
\author{Matthias Bode} 
	\address{Physikalisches Institut, Experimentelle Physik II, 
	Universit\"{a}t W\"{u}rzburg, Am Hubland, 97074 W\"{u}rzburg, Germany}	
	\address{Wilhelm Conrad R{\"o}ntgen-Center for Complex Material Systems (RCCM), 
	Universit\"{a}t W\"{u}rzburg, Am Hubland, 97074 W\"{u}rzburg, Germany}   
\date{\today}

\begin{abstract}
We have studied cleaning procedures of Nb(110) by verifying the surface quality with low-energy electron diffraction, 
Auger electron spectroscopy, and scanning tunneling microscopy and spectroscopy.  
Our results show that the formation of a surface-near impurity depletion zone is inhibited 
by the very high diffusivity of oxygen in the Nb host crystal 
which kicks in at annealing temperatures as low as a few hundred degree Celsius. 
Oxygen can be removed from the surface by heating the crystal up to  $T = 2400^\circ$C. 
Tunneling spectra measured on the clean Nb(110) surface exhibit a sharp conductance peak 
in the occupied states at an energy of about $-450$\,meV. 
Density functional theory calculations show that this peak is caused by a $d_{z^2}$ 
surface resonance band at the $\bar{\Gamma}$ point of the Brillouin zone
which provides a large density of states above the sample surface.  
The clean Nb(110) surface is superconducting with a gap width 
and a critical magnetic field strength in good agreement to the bulk value.  
In an external magnetic field we observe the Abrikosov lattice of flux quanta (vortices). 
Spatially resolved spectra show a zero-bias anomaly in the vortex core.  
\end{abstract}
\maketitle

\section{Introduction}

Recent proposals in topological quantum computation rely on the storage and manipulation 
of exotic quasiparticles, such as anyons.\cite{Kitaev2003, Nayak2008} 
For example, under special conditions zero-energy Majorana fermions were predicted 
to exist in topological superconductors, i.e., at the ends of one-dimensional nanowires 
or in vortices of two-dimensional films.\cite{Alicea2010, Beenakker2013, Ivanov2001}  
However, the existence of bulk materials which exhibit topological superconductivity is still elusive.  
An alternative approach focused on hybrid materials, e.g., 
by combining a strongly spin-orbit--coupled \textit{s}-wave superconductor 
with permanently ordered magnetic moments in its direct proximity. 
\cite{Nadj-Perge2013, Yazdani2014, Menard2017, Kim2018} 
Indeed, zero-bias conductance peaks consistent with expectations for Majorana fermions 
were experimentally observed at the ends of a self-assembled Fe chains on Pb(110)\cite{Yazdani2014} 
and also at edges of two-dimensional Co islands in the Pb/Co/Si(111) system.\cite{Menard2017} 
A more recent approach used atomic manipulation with an STM tip 
to create single-atom Fe chains on Re(0001).\cite{Kim2018} 

The experimental results of these studies illustrate that the choice of the substrate 
will play a decisive role towards an unambiguous demonstration of topological quantum computation.  
Namely, Pb has the advantage of a relatively high superconducting critical temperature $T_{\rm c} = 7.2$~K, 
but its relatively large lattice constant and low cohesion energy make it unsuitable for single-atom manipulation, 
severely limiting their potential for the creation of nano-engineered structures.  
Conversely, the much harder Re(0001) surface is a good platform for the creation of atomic-scale nanostructures, 
but Re exhibits a much lower $T_{\rm c}$ such that spectroscopic features are easily smeared out 
by thermal broadening at experimentally accessible temperatures.  
With these considerations in mind, Nb appears to be a promising material
since it combines a high cohesive energy with a sizable spin-orbit coupling 
and a wide superconducting gap ($2\Delta = 3.05$~meV and $T_{\textrm{c}} \approx 9.2$~K). 
Furthermore, it is a type-II superconductor which opens up the possibility 
of searching for Majorana states inside of vortices in an external magnetic field. 

The main drawback of Nb, however, is its affinity to oxygen (O) 
which leads to a surface that is notoriously difficult to prepare.  
The standard cleaning procedure reported in literature\cite{Haas1966,Pantel1977,Franchy1996,Suergers2001,Razinkin2010}
begins with cycles of ion sputtering and annealing at moderate temperatures up to about $1900^\circ$C.   
It has clearly been stated, however, that these initial sputter-annealing cycles---although 
they are suitable to remove carbon (C) and nitrogen (N) from the surface 
and reduce the surface-near concentration of O---are not suitable 
to clean the Nb(110) surface since a significant amount of O remains.  
These O-reconstructed surfaces have been extensively investigated 
by low-energy electron diffraction (LEED),\cite{Haas1966,Haas1967,Haas1968,Pantel1977,Franchy1996}
Auger electron spectroscopy (AES),\cite{Pantel1977,Franchy1996}
electron-energy loss spectroscopy (EELS),\cite{Franchy1996} 
and scanning tunneling microscopy and spectroscopy (STM/STS).\cite{Suergers2001,Razinkin2010}

There seems to be agreement that clean Nb(110) can only be obtained 
when heating the sample up to $T > 2200^\circ$C.\cite{Haas1966,Pantel1977,Franchy1996,Razinkin2010} 
In this case LEED experiments showed only diffraction maxima expected for clean Nb(110). 
However, LEED averages over a large area and is therefore not suitable 
to obtain a microscopic understanding of the details of the cleaning process.  
In this work we describe step-by-step the cleaning procedure of Nb(110). 
While the focus is clearly on STM, we also applied LEED and AES depth profiling 
to better understand the limitations of the various procedures.  
Furthermore, the electronic properties of the clean Nb(110) surface 
were investigated by scanning tunneling spectroscopy (STS) and density functional theory (DFT) calculations.
Our results confirm the existence of a surface resonance of $d_{z^2}$ orbital character 
in the local density of states (LDOS) which is energetically located about $450$\,meV below the Fermi level $E_{\rm F}$.
We image the Abrikosov lattice produced by an external magnetic field 
by mapping the differential conductance in and around the superconducting gap. 
We observe a strong conductance peak at the Fermi level in the vortex core, 
which we interpret as Caroli-de-Gennes-Matricon (CdGM) states. 

\section{Methods}
\subsection{Experimental setup and measurement procedures}

Investigations were performed with rectangular-shaped Nb(110) single crystals (7\,mm $\times$ 7\,mm $\times$ 1\,mm)
which were polished on one side with a nominal miscut of $0.1^{\circ}$.
They were fixed onto flag style tungsten sample plates by means of two tungsten wires. 
To prevent alloy formation at the crystal--sample holder interface 
they were separated by stacks consisting of a Nb and a W foil. 
After introduction into the UHV system, the crystal used for the STM investigations 
in Sect.\,\ref{sec:cleaning} was ion-sputtered for two hours at room temperature. 
This sample was initially heated several dozen times at $T = 1800^\circ$C.  %($T = 1200^\circ$~C).
The main contamination of Nb is oxygen which is dissolved in the bulk and segregates to the surface after each flash heating cycle. 
Indeed, our attempts to create an oxygen depletion layer at the surface through multiple cycles of ion sputtering 
and subsequent annealing at $T \approx 1800^\circ$C did not result in sufficiently clean surfaces. 

As we will point out in more detail in Sect.\,\ref{sec:cleaning} below the only way to get rid of oxygen 
is to flash heat the Nb crystal close to its melting point ($T_{\rm melt} = 2477^\circ$C). 
We use an $e$-beam heating stage to directly bombard the polished side of the Nb(110) surface with electrons. 
We measured the temperature with an optical pyrometer (Ircon UX-70P) through Kodial glass windows.
Since the viewport is located under an angle of $\approx 54^\circ$ with respect to the crystal's surface normal 
we corrected the measured temperature for the deviation from Lambert's cosine law.\cite{SupplMat}   
We gradually increased the heating temperature of the Nb(110) crystal. 
Each flash lasted for 30\,s, whereby the maximum surface temperature was typically reached after about 15\,s.
Depending on the maximum pressure in the UHV chamber during the high temperature flash it was repeated 2-4 times 
with the aim to ensure that the pressure stays below $p < 5 \times 10^{-8}$\,mbar. 

After each step we analysed the evolution of the Nb(110) surface by STM measurements 
performed with a home-built cryogenic scanning tunneling microscope which is equipped 
with a superconducting magnet with a maximum magnetic field of 12\,T perpendicular to the sample surface. 
The minimal achievable tip and sample temperature inside the STM amounts to $T = 1.2$\,K. 
All measurements were performed with electro-chemically etched W tips, 
which were preliminary tested on a Ag(111) surface. 
For spectroscopic measurements of the differential conductance a small modulation voltage $U_{\rm mod}$
with a frequency $f = 890$\,Hz was superimposed to the sample bias $U$ 
such that the frequency- and phase-selective d$I$/d$U$ signal can be detected at low noise by means of a lock-in amplifier.  
The resulting d$I$/d$U$ spectrum was verified by comparison 
with the numerically calculated d$I$/d$U$ signal from averaged $I(U)$ curves. 
Auger electron spectroscopy and low energy electron diffraction experiments 
were performed at room temperature with a retarding field four-grid electron optics. 

\subsection{Computational methods}
We used the plane wave-based \textsc{vasp}~\cite{vasp1,vasp2} code 
within the projector augmented-wave method (PAW)~\cite{blo,blo1} 
for our \emph{ab initio} density functional theory (DFT) calculations. 
The generalized gradient approximation (GGA) 
of Perdew-Burke-Ernzerhof (PBE)~\cite{PBE,PBEerr} was used for the exchange-correlation. 
We used the PBE functional as the PBE calculated lattice constant is 3.31\,{\AA}, 
in good agreement with the experimental bulk Nb lattice constant of 3.296\,{\AA}.~\cite{Haas2009} 
The energy cutoff for the plane wave basis set was set at 350\,eV.
	
The clean Nb(110) surface system was modelled using a symmetric slab consisting of 19 atomic layers of Nb. 
A 16\,{\AA}-thick vacuum layer was included in the direction normal to the surface 
to remove any spurious interactions between the repeating slabs. 
Structural relaxations via force optimization have been performed 
to obtain the surface interlayer distance for the three surface layers only. 
The positions of the other layers were kept constant. 
We found an average interlayer distance for the surface layers $\sim 2.28$\,{\AA} 
which is in a good agreement with the experiments. 
For structural relaxations, we have used $10^{-2}$\,eV/{\AA} force tolerance 
and a $\Gamma$-centred $5 \times 5 \times 1$ $k$-point mesh. 
2025 $k_\parallel$-points in the full two-dimensional Brillouin zone were used 
for the calculations of the band structure and other electronic properties. 
% For the calculation of simulated STM images, we used  % the extension of 
% the Tersoff-Hamann model where the tunneling current is proportional to the local density of states (LDOS), 
% $I(\bm{R}_{\rm T}) \propto n_{\mathrm{S}}(\bm{R}_{\rm T},E_{\rm F})$.~\cite{Tersoff1983,Tersoff1985}

%%%%  RESULTS   %%%%
\section{Surface structure}

\subsection{Closed NbO films on Nb(110)} \label{subsec:NbO_films}
\begin{figure}[t]
	\includegraphics[width=0.7\columnwidth]{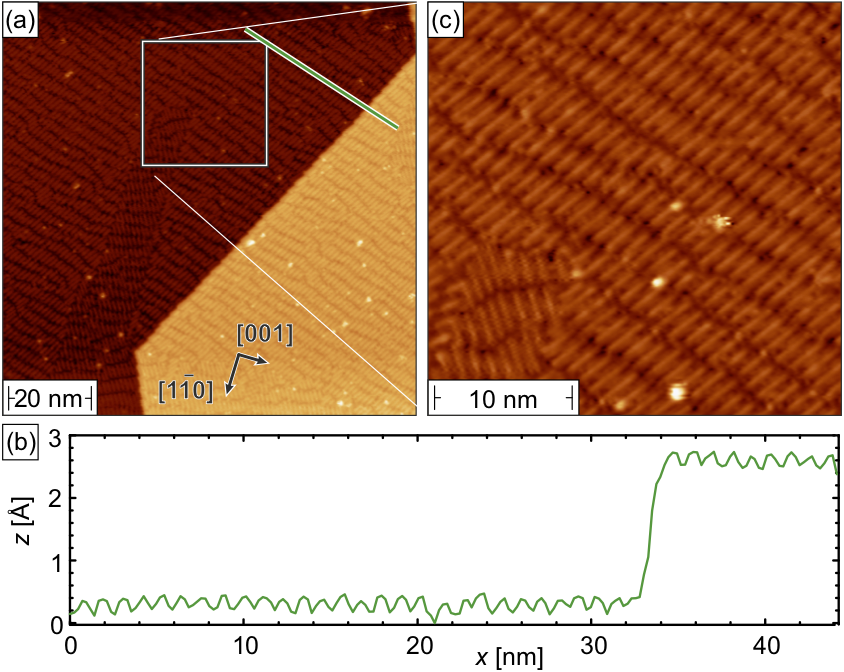}
	\caption{(a) Overview of a Nb(110) surface after heating at $T = 1800^\circ$C. 
	The surface exhibits periodically arranged stripes which are oriented in two different spatial directions, 
	fully covered with an oxygen-induced reconstruction. 
	(b) Line profile measured along the green line in (a). 
	(c) Detailed view of the area marked by a box in (a). 
	Stabilization parameters: $U =-100$\,mV, $I =100$\,pA.}
	\label{fig:Nb_1st_topo}
\end{figure}
In a first step, the Nb(110) sample was heated to a maximum temperature $T = 1800^\circ$C. 
Fig.\ \ref{fig:Nb_1st_topo}(a) displays an overview image (scan area: 100\,nm $\times$ 100\,nm) of the resulting sample surface.  
One can recognize two terraces, which are separated by a step edge of monatomic height. 
The upper part and lower part of the step edge are oriented under an angle of about $120^{\circ}$, 
a value which significantly exceeds the angle expected between two equivalent $\langle 111\rangle$ directions 
within a bcc(110) surface plane, $[1\bar{1}1]$ and $[\bar{1}11]$, of $109.5^{\circ}$.  
More careful inspection shows that both terraces are not smooth but instead exhibit 
a sawtooth pattern with short segments oriented along $\langle 111\rangle$ directions.   

A line profile drawn along the green line is plotted in Fig.\ \ref{fig:Nb_1st_topo}(b).
It shows a step height of $(2.3 \pm 0.1)$\,{\AA}, in a good agreement 
with the corresponding bulk value $a_{\textrm{Nb(110)}}^{\textrm{bulk}} =  2.36$\,{\AA}. 
Closer inspection reveals that the surface is not perfectly flat but exhibits periodically arranged stripes 
with a periodicity of $(1.15 \pm 0.05)$\,nm and a corrugation of about 20\,pm at $U = -100$\,mV.   
Fig.\ \ref{fig:Nb_1st_topo}(c) shows a higher resolution close-up of the area marked by a white box in panel (a). 
Now we clearly recognize stripes that are oriented along $\langle 111\rangle$ directions and about $(3.3 \pm 0.3)$\,nm long. 
Since the overall appearance observed by us is in good agreement with published STM data 
obtained on oxygen-reconstructed Nb(110) surfaces \cite{Suergers2001,Razinkin2010} 
we can safely assume that the origin of the reconstruction is surface-segregated oxygen. 
For more details on the atomic arrangement and crystallography of NbO films on Nb(110) 
the interested reader should refer to Refs.\,\onlinecite{Suergers2001} and \onlinecite{Razinkin2010}.

\subsection{Attempts of cleaning Nb(110) by sputter--anneal cycles}
\begin{figure}[b]
	\includegraphics[width=0.6\columnwidth]{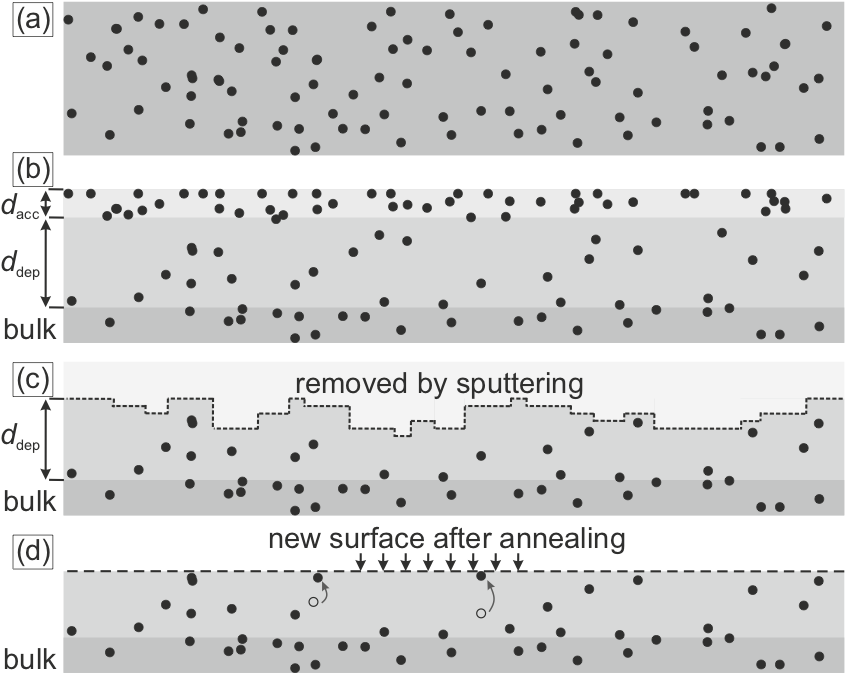}
	\caption{Schematic representation of the Nb cleaning process.
		(a)~Homogeneously distributed impurities (black dots) in the bulk-terminated Nb crystal (gray).
		(b)~Annealing of Nb leads to surface segregation of O impurities,
		i.e.\ an enhanced/reduced concentration in the accumulation/depletion zone (thickness $d_{\rm acc}$/$d_{\rm dep}$).
		(c)~Sputtering removes the accumulation zone but also roughens the surface.
		(d)~Annealing flattens the surface but may again result in a surface segregation. }
	\label{fig:AnnScheme}
\end{figure}
In an attempt to apply cleaning procedures similar to those of other $4d$ and $5d$ transition metal elements, 
such as Mo or W,\cite{KGO2000,BKB2007,ZPZ2010} we prepared the Nb(110) surfaces with numerous sputter and annealing cycles.  
Fig.~\ref{fig:AnnScheme} shows a schematic representation of the mechanism. 
The native surface which is usually created by cutting and polishing a bulk crystal exhibits a homogeneous density of impurities.  
In the case of Nb the main contaminants will be O and C [Fig.~\ref{fig:AnnScheme}(a)]. 
As the Nb crystal is heated the O impurities segregate at the surface [Fig.~\ref{fig:AnnScheme}(b)]. 
The thickness of the accumulation zone, where the density of impurities is enhanced 
as compared to the native crystal, is denoted by $d_{\rm acc}$. 
Since the total number of impurities has to be conserved the O surface segregation 
inevitably leads to a sub-surface depletion zone underneath the accumulation zone. 
The thickness of the depletion zone $d_{\rm dep}$ will depend on the O impurity concentration of the particular Nb crystal, 
on the surface and bulk diffusion constants of O in Nb, and on the actual annealing temperature. 
The surface-segregated O is removed by sputtering at room temperature 
which inevitably roughens the surface [Fig.~\ref{fig:AnnScheme}(c)]. 
This sputter-induced surface roughness can be removed by a final annealing step, 
as schematically presented in Fig.~\ref{fig:AnnScheme}(d). 
We have to keep in mind however, that this final annealing step may again lead 
to the unwanted segregation of bulk impurities to the surface, 
symbolized by curved arrows ($\curvearrowright$) in Fig.~\ref{fig:AnnScheme}(d).

\begin{figure}[b]   %%%%%%%%%%%%%%%%%%%%%%%%%%%%%%%%%%%%%%%%%%%%%%
	\begin{minipage}[t]{0.65\textwidth} 
		\includegraphics[width=1\columnwidth]{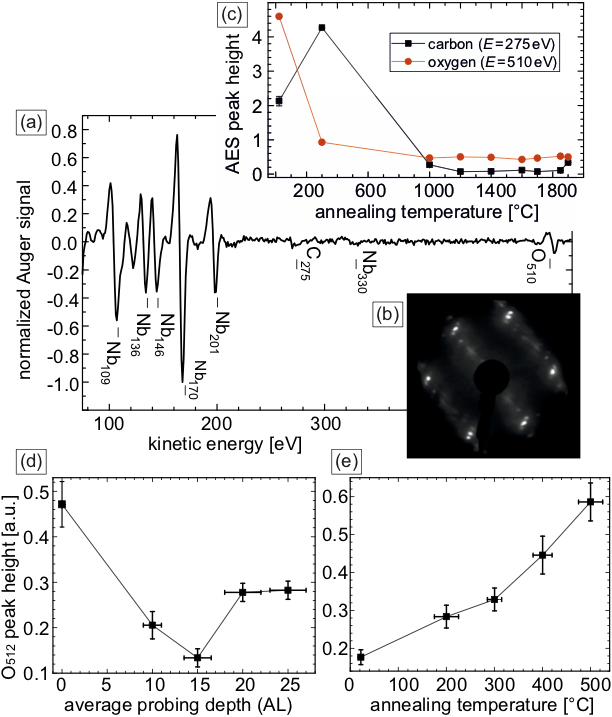}%
	\end{minipage}
	\hfill
	\begin{minipage}[b]{0.3\textwidth}
		\caption{(a)~Auger spectrum and (b) LEED pattern of Nb(110) after heating at $T = 1600^{\circ}$C.
                (c)~Annealing temperature-dependent Auger intensities of C- and O-related peaks.
                (d)~Depth profile of the O-related Auger intensity obtained after sputter removal.  
                (e)~Intensity of the O peak after sputtering and subsequent annealing. } 
		 \label{fig:DeplLayer}
	\vspace{2cm}
	\end{minipage}	
\end{figure}    %%%%%%%%%%%%%%%%%%%%%%%%%%%%%%%%%%%%%%%%%%%%%%
In order to investigate whether the removal of surface-segregated O by sputtering and subsequent surface annealing 
represents a valid procedure to obtain clean Nb(110) surfaces we started with a newly received single crystal.   
This sample was heated at increasing temperatures for 1\,min up to maximum $T = 1900^{\circ}$C. 
After each step we took an Auger spectrum and inspected the LEED pattern.  
The data sets obtained after heating at $T = 1600^{\circ}$C are shown in Fig.\ \ref{fig:DeplLayer}(a) and (b), respectively. 
Fig.\ \ref{fig:DeplLayer}(c) shows the annealing temperature-dependent intensity 
of the C- and O-related peaks after normalization to the Nb signal.  
Whereas the O-related signal monotonously decreases with increasing heating temperature, 
the C peak increases at the initially heating step ($T = 200^{\circ}$C) but then rapidly decreases. 
Upon heating to $T = 1200^{\circ}$C the C- and the O-related signal 
have decreased to about 10\% of their starting values.  
However, within the error of our analysis no further improvement 
can be achieved when heating to $1200^{\circ}$C $< T \le 1900^{\circ}$.  

This heating leads to the formation of an O depletion layer directly underneath or in close proximity to the surface.   
To show this we successively removed material from the surface by Ar$^+$ sputtering 
at an ion energy $E = 1$\,keV and a sample current $I_{\rm ion} = 2\,\mu$A at a rate of $(0.30 \pm 0.02)$\,atomic layers (AL)/min. After each sputter step we analyzed the height of the O-related Auger peak. 
The resulting depth profile is plotted in Fig.\,\ref{fig:DeplLayer}(d).  
The data reveal that the O concentration is enhanced directly at the surface by about 50\% 
as compared to the bulk value represented by the data points at an average probing depth $d_{\rm av} \ge 20$\,AL.  
A minimum with an O concentration as low as half the bulk value is observed at $d_{\rm av} = 15$\,AL, 
representative for the depletion zone we were looking for.  

Unfortunately, this depletion zone is not stable against further annealing, as shown in Fig.\,\ref{fig:DeplLayer}(e).
These data have been achieved on a sample where the depletion zone has been exposed to the surface 
by sputtering, analogous to the data point at $d_{\rm av} = 15$\,AL in Fig.\,\ref{fig:DeplLayer}(d).  
Subsequently, this sample was annealed at successively increasing temperature for 4\,min at each heating step.  
Our data reveals that even a temperature $T_{\rm ann} = 200^{\circ}$C, which is much lower 
than what would realistically be needed for a proper annealing of the sputter-roughened Nb surface, 
results in a significant segregation of O at the surface.  
Therefore, we conclude that sputter and annealing cycles 
do not represent a suitable cleaning method for Nb(110) surfaces.  

\subsection{Cleaning Nb(110) by heating} \label{sec:cleaning}

\begin{figure*}
	\includegraphics[width=\textwidth]{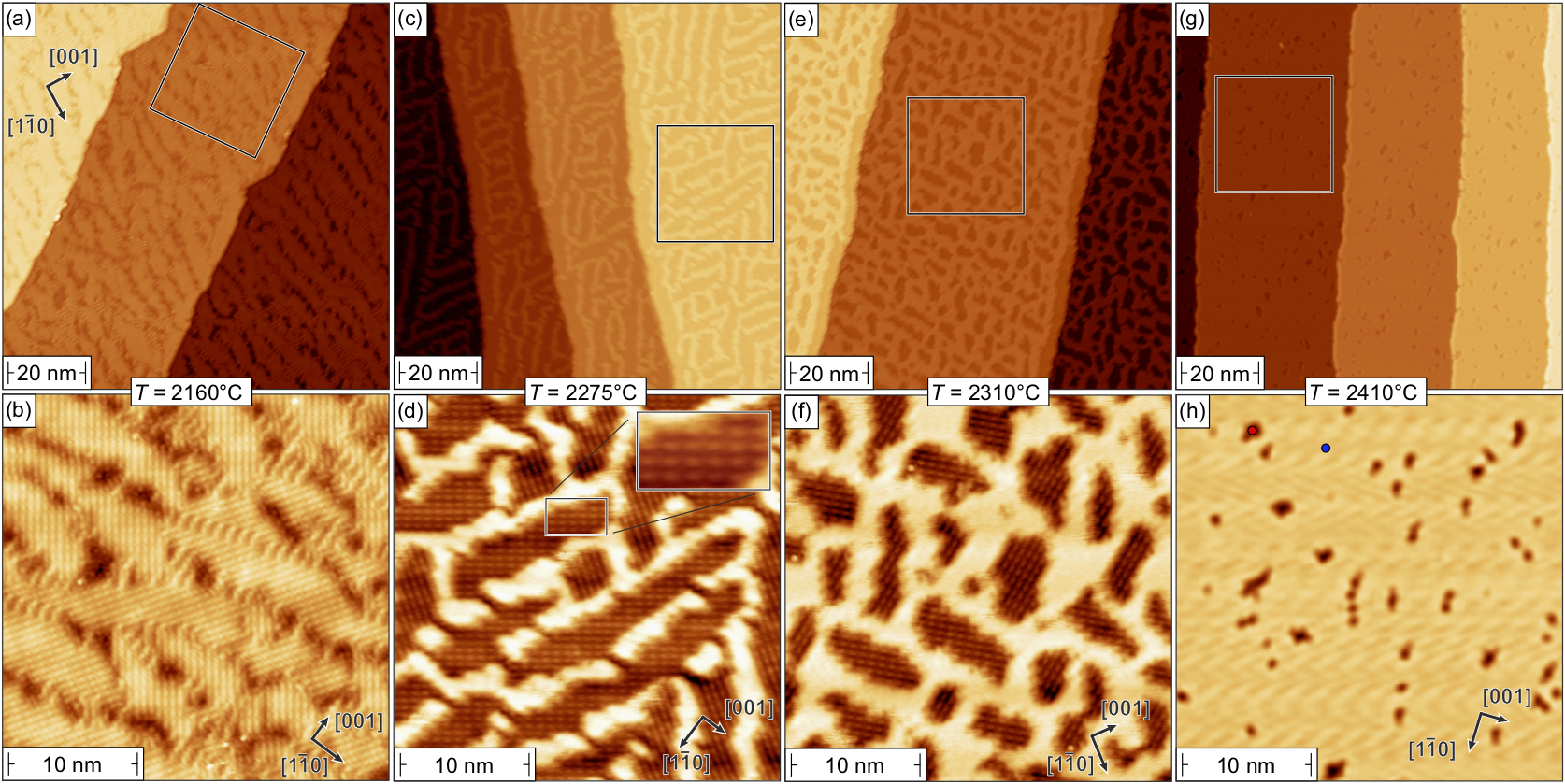}
	\caption{Topography of Nb(110) after flash heating at different temperatures. 
		In each case an overview image (scan area: 100\,nm $\times$ 100\,nm) 
		and a higher resolution scan (25\,nm $\times$ 25\,nm) on an atomically flat terrace are shown.
		(a),(b) Nb(110) after heating cycles at  $T_{\textrm{surf}} = 2160^\circ \textrm{C}$, 
		showing an oxygen reconstruction and trench-like depressions. 
		(c),(d) Heating cycles at $T_{\textrm{surf}} = 2275^\circ \textrm{C}$ 
		result in a surface with unreconstructed Nb(110) covering about 25\% of the total area. 
		The inset shows a zoom-in of the oxygen reconstruction also observed in (b).
		(e),(f) The surface fraction of unreconstructed Nb(110) increases to 60\% 
		after heating cycles at $T_{\textrm{surf}} = 2310^\circ \textrm{C}$. 
		(g),(h) Best surface quality obtained for Nb(110) after heating cycles 
		at $T_{\textrm{surf}}=2410^\circ \textrm{C}$.   This surface is about 95\% clean. 
		Stabilization parameters: (a)-(h) $U = -1$\,V, $I = 100$\,pA.}
	\label{fig:Nb_2nd_topo}
\end{figure*}

In the following, we attempted to clean the Nb(110) surface 
only by heating the surface by electron-beam bombardment at higher temperatures.  
As we will point out below this resulted in a very high surface quality 
with an impurity concentration well below 5\% and electronic properties characteristic for clean Nb(110).  
Upon heating at a temperature $T_{\textrm{surf}} = 2160^\circ \textrm{C}$, 
the surface still remains largely covered with oxygen.  
While the overview image of Fig.\,\ref{fig:Nb_2nd_topo}(a) only indicates the presence of some dark spots, 
we can clearly recognize that the surface is reconstructed 
in the higher resolution scan displayed in Fig.\,\ref{fig:Nb_2nd_topo}(b).
Also on this surface domains with two stripe orientations exist, which are separated by trench-like depressions. 
We would like to note, however, that the type of reconstruction differs 
from what has been discussed in Sect.\,\ref{subsec:NbO_films} 
and Refs.\,\onlinecite{Suergers2001} and \onlinecite{Razinkin2010}, 
as can be seen by close inspection of Fig.\,\ref{fig:Nb_2nd_topo}(b). 
A detailed analysis of this oxygen-induced reconstruction will be given below. 

With further heating ($T_{\textrm{surf}} = 2280^\circ \textrm{C}$), the oxygen-reconstructed domains become smaller, 
thereby opening up space for narrow strips which appear significantly higher 
in constant-current STM images [Fig.\,\ref{fig:Nb_2nd_topo}(c) and (d)].  
While the precise height difference between reconstructed and higher surface areas 
depends on the particular scan parameters (bias voltage), the latter are generally characterized by their smooth surface. 
The high density of states and the unreconstructed surface indicate that these surface areas are clean Nb(110). 
When the maximum temperature is raised even closer to the melting point of Nb, 
the oxygen-reconstructed regions become progressively smaller. 
For example, at $T_{\textrm{surf}} = 2310^\circ \textrm{C}$ unreconstructed Nb(110) 
covers already about 60\% of the total surface area  [Fig.\,\ref{fig:Nb_2nd_topo}(e) and (f)].  
Eventually, at $T_{\textrm{surf}} = 2410^\circ \textrm{C}$ the surface consists to $\approx 95$\% of pure Nb(110) and only small dark areas indicate that some oxygen still remains on the surface [Fig.\,\ref{fig:Nb_2nd_topo}(g) and (h)]. 
We would like to admit, however, that during the final heating cycle some parts of our crystal were melted, 
presumably because of a small temperature gradient along the sample which accounts for a few tens of degrees.
We also observed various amounts of hydrogen contamination on Nb(110) at $T = 4.2\textrm{K}$,  
strongly dependent on the level of hydrogen in the UHV chamber and---even more important---in the cryostat. 
Hydrogen can be seen only at low bias voltages $\vert U \vert \leq 0.2$\,V 
and can be removed from the scan area by the STM tip, probably due to electron-stimulated desorption.\cite{SupplMat}
 
\subsection{Properties of the dilute NbO phase} \label{sec:dil_NbO}

Also at oxygen concentrations which are lower than what has been discussed above 
for closed NbO films in Sect.\,\ref{subsec:NbO_films}, we observe two domains 
with row-like structures that are rotated by $(110 \pm 2)^\circ$ relative to each other. 
However, the specific parameters of this reconstruction 
significantly differ from the values determined for NbO films.  
For example, the inter-row distance amounts to $(5.7 \pm 0.1)$\,{\AA} only. 
Furthermore, close inspection, e.g., of the reconstructed regions in Fig.\,\ref{fig:Nb_2nd_topo}(d) 
which is shown at higher magnification in the inset, reveals that these lines 
consist of short protrusions with a periodicity of $(8.3 \pm 0.6)$\,{\AA}.  
\begin{figure}[h]
	\includegraphics[width=0.63\columnwidth]{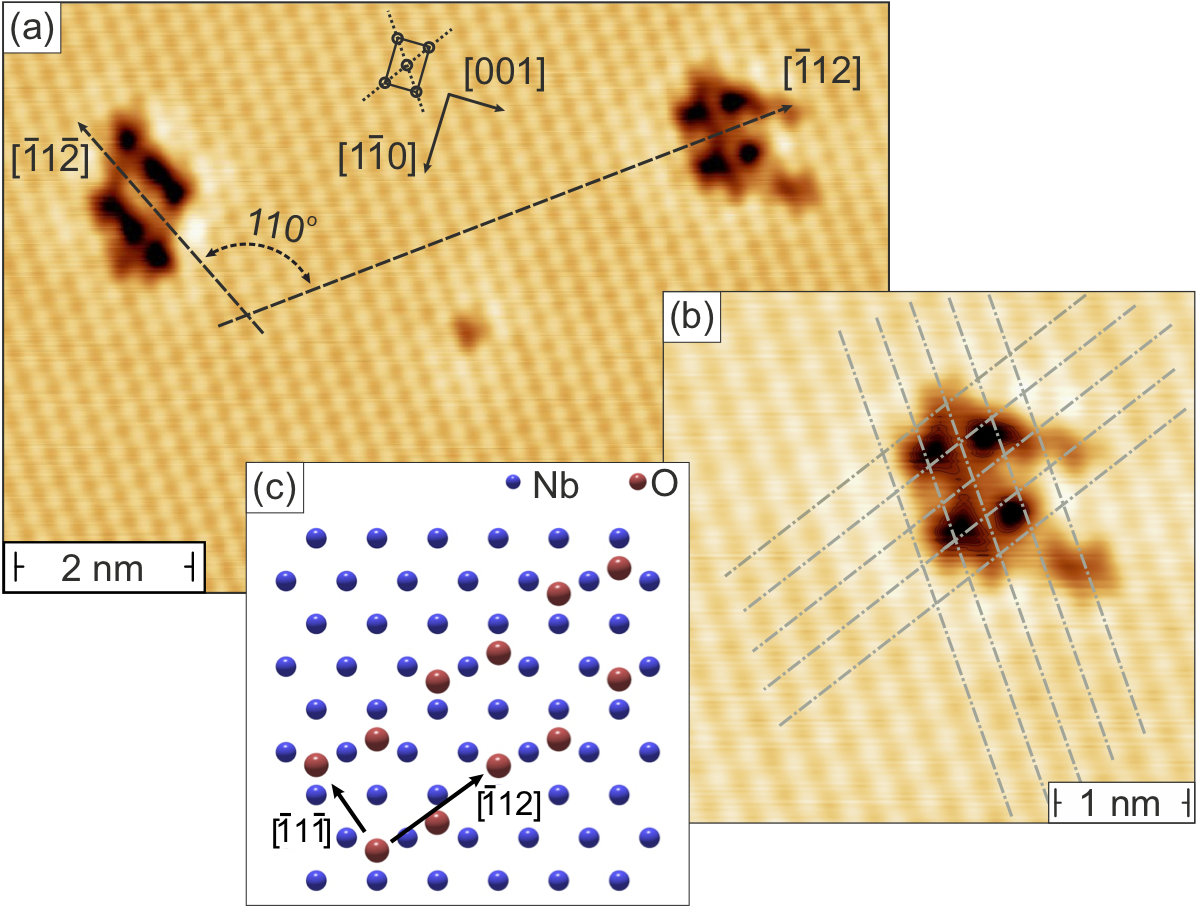}
	\caption{(a) STM image of oxygen atoms (dark) with atomic resolution on clean Nb(110). 
		(b) Zoomed-in image of the upper right part of (a) 
		superimposed with gray dot-dashed lines that interpolate 
		the atomic corrugation maxima of Nb(110). 
		Oxygen sits in a threefold-coordinated hollow site of the Nb lattice. 
		Stabilization parameters:  $U = -0.1$\,V, $I = 1$\,nA.
		(c) Hard ball model showing the possible atomic structure of this oxygen reconstruction.}
	\label{fig:O_adsorp_pos}
\end{figure}

Whereas we could not obtain atomic resolution images to determine the exact crystalline structure 
of this second reconstruction on samples with extended oxygen-induced domains, 
the data measured on a Nb(110) sample prepared at $T_{\textrm{surf}} = 2410^\circ \textrm{C}$ 
which are shown in Fig.\,\ref{fig:O_adsorp_pos}(a) allow for the identification of the oxygen adsorption site.  
It shows rows of depressions (dark contrast) oriented along the $[\bar{1}12]$ 
and the $[\bar{1}1\bar{2}]$ directions, i.e., under an angle of $110^\circ$. 
Details of our data analysis with an interpolation of the Nb atomic-scale corrugation maxima 
indicated by gray dot-dashed lines are shown in Fig.\,\ref{fig:O_adsorp_pos}(b).   
It reveals that the depressions are preferentially located at threefold-coordinated hollow sites. 
This adsorption site with an average Nb--O bond length of $2.09$\,{\AA} 
was indeed predicted by first-principles calculations.\cite{Tafen2013} 
The observed inter-row distance as well as the periodicity along the stripes
can consistently be explained with the toy-model presented in Fig.\,\ref{fig:O_adsorp_pos}(c).

\section{Electronic properties}
\subsection{Band structure}

\begin{figure}[t]   %%%%%%%%%%%%%%%%%%%%%%%%%%%%%%%%%%%%%%%%%%%%%%
	\begin{minipage}[t]{0.45\textwidth} 
		\includegraphics[width=\columnwidth]{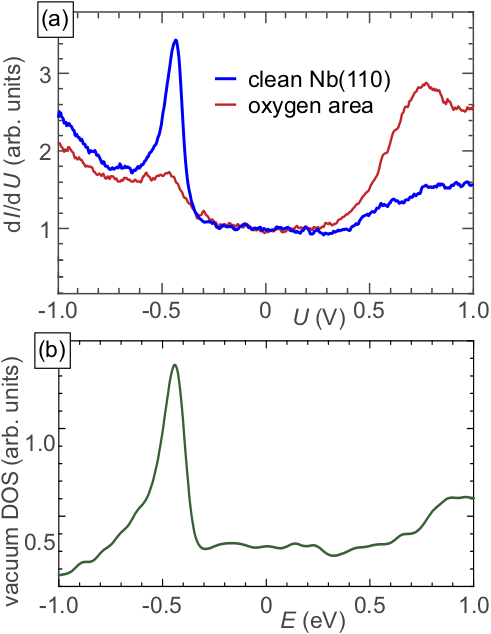}
	\end{minipage}
	\hfill
	\begin{minipage}[b]{0.45\textwidth}
		\caption{(a) Tunneling spectroscopy data of the clean (blue) and oxygen-reconstructed (red) Nb(110) surface. 
			Stabilization parameters:  $U = -1$\,V, $I = 0.1$\,nA, $T = 1.6$\,K
			(b) Calculated vacuum density of states of the clean Nb(110) surface 
			at a distance of $z = 4$ {\AA} above the surface.} 
		 \label{fig:Nb_STS_data}
	\vspace{2.5cm}
	\end{minipage}	
\end{figure}

Figure\,\ref{fig:Nb_STS_data}(a) shows tunneling spectroscopy data 
taken over the clean Nb(110) surface and over an oxygen-reconstructed area. 
The exact locations where these STS data were measured are marked in Fig.\,\ref{fig:Nb_2nd_topo}(h) 
with a blue dot for the clean surface and a red dot over the dark oxygen area. 
A sharp peak at $U \approx -0.45$\,V can be recognized for clean Nb 
which is strongly suppressed above oxygen-reconstructed areas. 
The local density of states (DOS) of the clean Nb(110) as calculated in the vacuum 
at $z = 4$\,{\AA} above the surface is shown in Fig.\,\ref{fig:Nb_STS_data}(b).
It is in excellent agreement with the experimental data 
not only with respect to the peak position but also regarding the peak shape.  
In addition, the experimental and the theoretical spectrum 
exhibit a relatively flat signal in the range $-0.3$\,eV\,$\le E \le +0.5$\,eV,   
which slightly rises as we proceed further into the unoccupied states, i.e., $E > +0.5$\,eV.
The only qualitative difference between experiment and theory can be found at $E < -0.6$\,eV, 
where the experimental data show a rising $\mathrm{d}I/\mathrm{d}U$ signal whereas the calculated DOS decreases.  
This difference could arise due to the change of the tunneling barrier at large bias voltages, 
which leads to an increase in $\mathrm{d}I/\mathrm{d}U$ signal and it is not taken into account by use the Tersoff-Hamann model, or it could come from tip states which often dominate tunneling spectra at large negative bias voltages.\cite{Ukr1996}
    %%%%%%%%%%%%%%%%%%%%%%%%%%%%%%%%%%%%%%%%%%%%%%

\begin{figure}[t]   %%%%%%%%%%%%%%%%%%%%%%%%%%%%%%%%%%%%%%%%%%%%%%
	\begin{minipage}[t]{0.55\textwidth} 
		\includegraphics[width=\columnwidth]{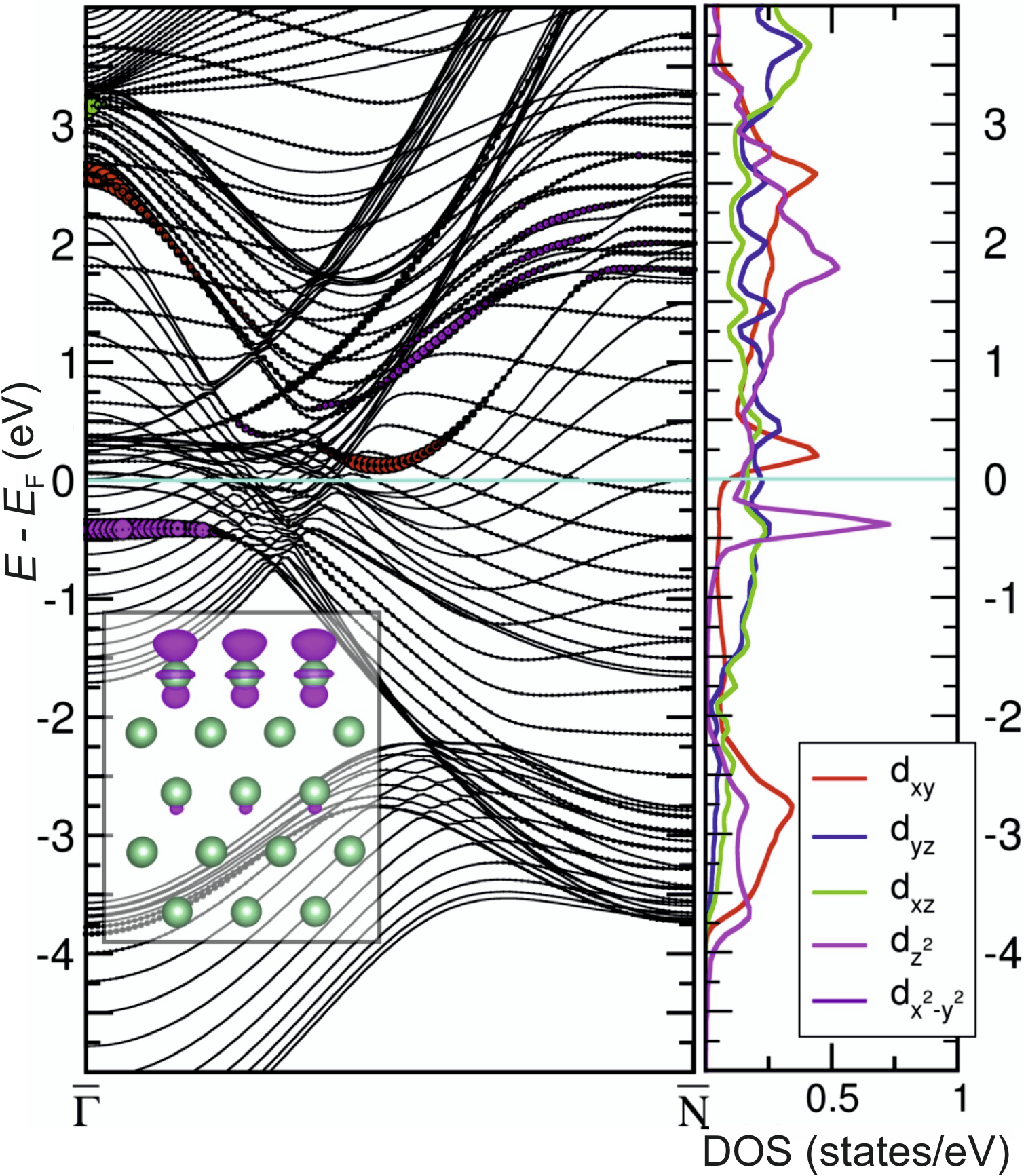}%
	\end{minipage}
	\hfill
	\begin{minipage}[b]{0.4\textwidth}
		\caption{Calculated band structure of clean Nb(110) 
		along the $\overline{\Gamma}\overline{\mathrm{N}}$ direction of the Brillouin zone. 
		States marked with colored circles are the different $d$ orbitals localized at the surface atoms. 
		Corresponding local density of states (LDOS) of $d$ orbitals for the surface atom are shown on the right panel. 
		The inset shows a side view of the atomic lattice (green) 
		and the charge density of the $d_{z^2}$ state at $E - E_{\rm F} = -0.45$\,eV (purple). } 
		 \label{fig:Nb_bandstructure}
	\vspace{1cm}
	\end{minipage}	
\end{figure}    %%%%%%%%%%%%%%%%%%%%%%%%%%%%%%%%%%%%%%%%%%%%%
In order to obtain the band dispersion of Nb(110) and understand the origin 
of the sharp feature observed in theoretical and experimental tunneling spectra, 
we have investigated the band structure and the site-projected local density of states of the surface atoms. 
Fig.\,\ref{fig:Nb_bandstructure} shows the calculated band structure of clean Nb(110) 
along the $\overline{\Gamma \mathrm{N}}$ direction of the Brillouin zone. 
One can clearly observe a nearly flat surface resonance band at about $E - E_{\rm F} = -0.45$\,eV. 
The downward dispersion of this band leads to the characteristic shape 
of the peak in the vacuum LDOS that drops sharply towards higher energies. 
The orbital-decomposed local DOS of the surface atom 
and energy-resolved local charge density analysis clearly identifies a $4 d_{z^2}$ state 
as the origin of the sharp STS peak as well as the peak observed in vacuum LDOS. 
This band disperses from $\overline{\Gamma}$ to $\overline{\mathrm{N}}$. 
Our band structure matches quite well with previously published results.\cite{Smith1981}
A comparison of the total density of states in the bulk and above the surface can be found in Ref.\,\onlinecite{SupplMat}. 

%According to DFT calculation \cite{}, the sharp peak originates from an occupied $d_z$-like surface state 
%which lies about $-0.4$\,eV below the Fermi level (see Suppl. Mat.). 
%A comparison with computed band structure \cite{Smith1981}, 
%it was showed that this maximum comes from the $4d \Sigma_1$ band, 
%which disperses from $\Gamma$ to N, and takes a minimum just below the Fermi level. 
%This bulk band is extended to the surface in clean Nb, but attenuated there in case of contamination.

\subsection{Superconductivity}
\begin{figure}[b]
	\includegraphics[width=0.8
\columnwidth]{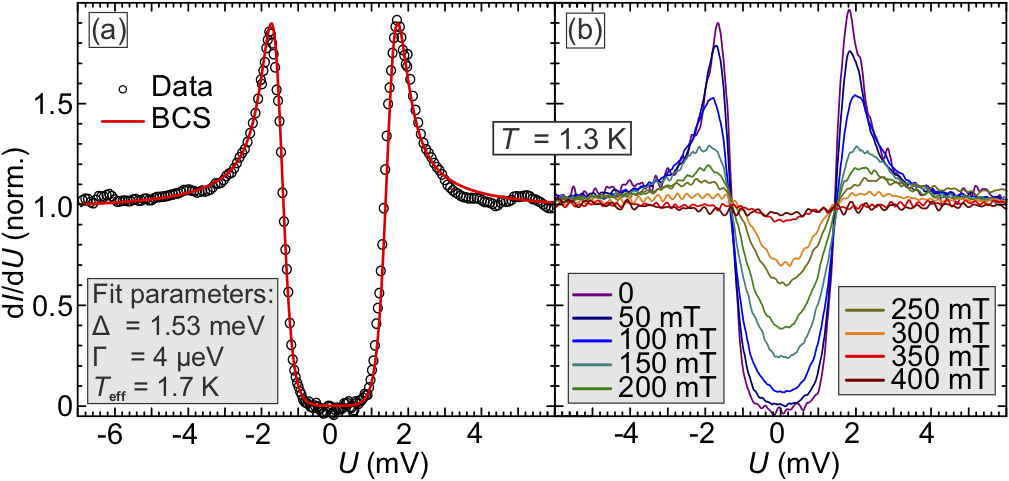}
	\caption{(a) Normalized spectroscopy of superconducting energy gap 
	as measured on clean Nb(110) at a temperature $T = 1.3$\,K. 
	The d$I$/d$U$ signal was fitted (red) by BCS theory; best fit parameters are specified. 
	(b) Normalized spectroscopy at different external magnetic fields. 
	As the field increases, the superconducting band gap weakens until it complete disappears at about 400\,mT. 
	Stabilization parameters: $U = -10$\,mV, $I = 400$\,pA, $U_{\textrm{mod}} = 0.1$\,mV.}
	\label{fig:Nb(110)_SCgap}	
\end{figure}
Figure\,\ref{fig:Nb(110)_SCgap}(a) shows an STS measurement 
of a superconducting gap as observed on the clean Nb(110) at a temperature $T = 1.3$\,K. 
A pronounced U-shaped superconducting energy gap can be seen, 
where the d$I$/d$U$ signal within the energy gap drops to zero. 
In order to fit the superconducting gap we employed the Dynes approach \cite{Dynes1978} with $\Delta = 1.53$\,meV and a quasiparticle lifetime broadening $\Gamma = 4$\,$\mu$V. 

\begin{figure}[b]
	\includegraphics[width=0.8\columnwidth]{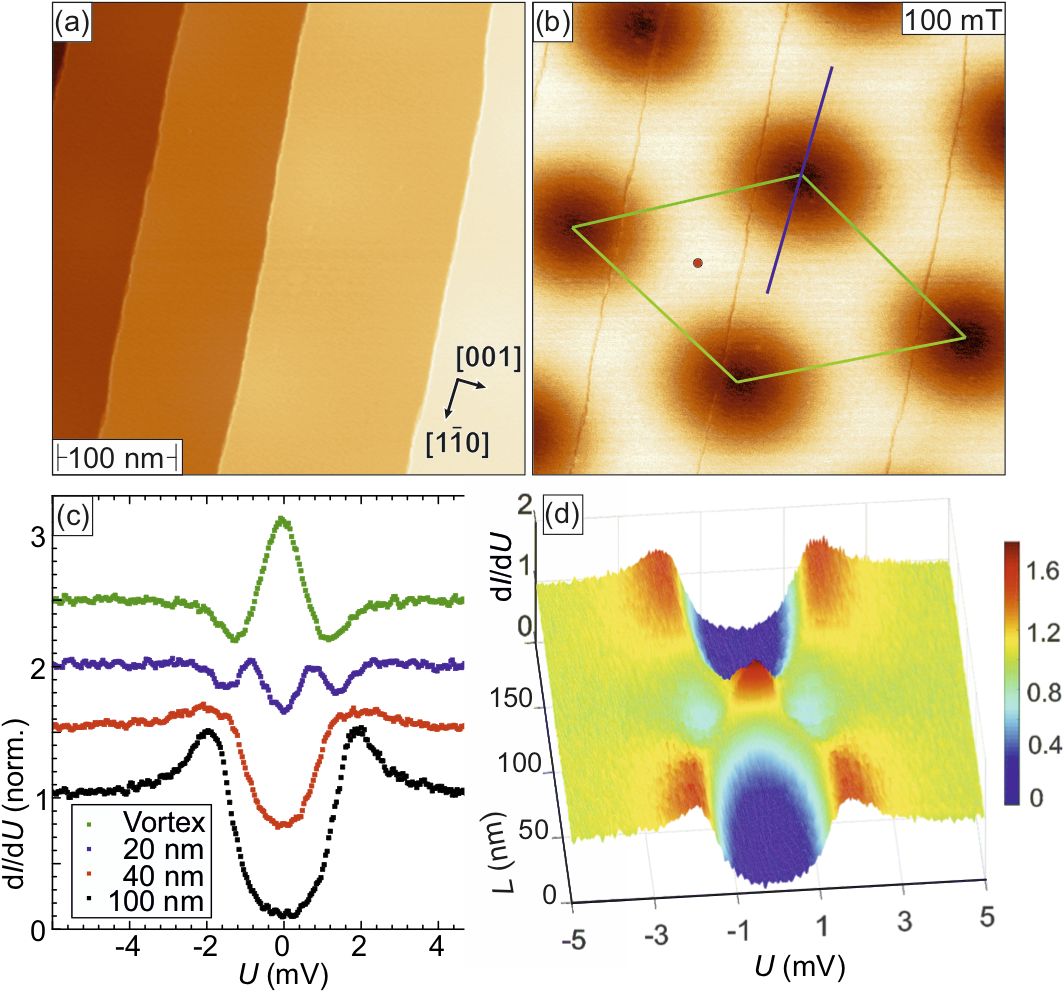}
	\caption{(a) Topography of a clean Nb(110) surface and (b) the simultaneously obtained 
		differential conductivity measured at a magnetic field 100\,mT. 
		Imaging parameter are $U = -1.8$\,mV, $I = 30$\,pA, $U_{\textrm{mod}} = 0.1$\,mV.
		The formation of a Abrikosov lattice can clearly be recognized. 
		(c) Tunneling spectra measured at various distances from the vortex core. 
		Whereas a well expressed superconducting energy gap 
		is observed at large distances ($\ge 100$\,nm, black curve), 
		a zero bias conductance peak develops inside the vortex core (green). 
		(d) Normalized differential conductivity as measured along the 200\,nm long 
		blue line in (b) as a function of vortex core distance. 
		Stabilization parameter: $U_{\textrm{set}} = -10$\,mV, $I_{\textrm{set}} = 200$\,pA, 
		$V_{\textrm{mod}} = 0.1$\,mV.  All data were measured at $T = 1.3$\,K. }
	\label{fig:Nb(110)_vortex}	
\end{figure}

The magnetic field dependence of the superconducting gap on clean Nb(110) is shown in Fig.\,\ref{fig:Nb(110)_SCgap}(b). 
Measurements were performed at different external magnetic fields, 
whereby it was ensured that the measurement took place as far away as possible from the surrounding vortices
[red dot in Fig.\,\ref{fig:Nb(110)_vortex}(b) and similar position for other magnetic fields]. 
With increasing magnetic field, the energy gap becomes weaker until it completely disappears at about 400\,mT. 
Accordingly, the upper critical field $B_{\textrm{c}}$ of the Nb(110) surface in our measurements 
is between 350\,mT and 400\,mT at a nominal sample temperature $T = 1.3$\,K. 
This result is in good agreement with values for Nb reported in literature.\cite{Finnemore1966, Karasik1970}

In a next step, we examined the superconductivity of clean Nb(110) 
in an external magnetic field with high spatial resolution. 
When a magnetic field is applied to a type-II superconductor, the Abrikosov vortex lattice 
with a quantized flux of $\Phi_0 = h/2e = 2.07 \times 10^{-15}$\,Wb per vortex is formed. 
Fig.\,\ref{fig:Nb(110)_vortex}(a) shows an overview image (400\,nm $\times$ 400\,nm) 
of a surface area which exhibits five atomically flat terraces separated by single-atomic step edges. 
Fig.\,\ref{fig:Nb(110)_vortex}(b) displays the differential conductivity map obtained simultaneously 
with the topography data in (a) at a bias voltage $U = -1.8$\,mV in an external magnetic field $B = 100$\,mT. 
At this bias voltage we tunnel from occupied sample states which are energetically located 
approximately at the maxima of the tunneling differential conductance at the edge of the superconducting energy gap. 
The contrast between superconducting areas (bright; high d$I/$d$U$ signal) and vortices (dark; low d$I/$d$U$ signal)
originates from the different density of states at the edge of the superconducting gap [also see panel (c) below].
We observed several round-shaped vortices with a diameter of $45 \pm 5$\,nm, 
in good agreement with the Nb coherence length $\xi$. 
The vortices are arranged in a triangular lattice with a nearest neighbour distance $185 \pm 10$\,nm, 
exceeding the expected unit cell parameter for the triangular Abrikosov lattice 
$a \approx \sqrt{\Phi_{0}/{B}} \approx 140$\,nm for a magnetic field $B = 100$\,mT. 
A similar discrepancy between the theoretically expected and experimentally observed 
vortex lattice parameter was also found at a higher magnetic field. 
It may be caused by the pinning of vortices in defect-rich regions of the Nb(110) crystal, 
possible at the crystal edges where heating-induced damages and melted spots first occur. 
Figure\,\ref{fig:Nb(110)_vortex}(c) shows differential conductance d$I/$d$U$ spectra taken at four points 
located at different distances from the center of the vortex core, i.e., 100\,nm, 40\,nm, 20\,nm, and directly in the vortex center. 
Directly in the vortex core we observe a zero-bias conductance peak (green curve)
which splits into two symmetrical maxima at 20\,nm (blue).   
At a distance of 40\,nm to the vortex core (red curve) the peaks 
at the edge of the gap are still weaker and the gap width is reduced
as compared to the spectrum recorded without any field [cf.\ Fig.\,\ref{fig:Nb(110)_SCgap}(a)].  
Only at distances much larger than $\xi$ (100\,nm, black curve), 
the spectrum closely resembles zero field data.

In order to better understand the spatial variations of these features, 
we took d$I/$d$U$ spectra taken across a superconducting vortex 
along the 200\,nm long line with a 4\,nm increment. 
The resulting data are plotted as waterfall plot in Fig.\,\ref{fig:Nb(110)_vortex}(d). 
The d$I$/d$U$ signal appears at the vertical scale and is additionally color-coded. 
At large distance from the vortex core a deep gap with pronounced side peaks can clearly be recognized. 
Both characteristics of superconductivity weaken towards the center 
until they essentially disappear at approximately $\pm 20$\,nm. 
There two weaker maxima split off from the edges of the former energy gap 
and converge into an X-shaped feature visible in yellow color in Fig.\,\ref{fig:Nb(110)_vortex}(d).
The crossing point is dominated by a strong maximum (red) marking the vortex core.
This peak in the d$I/$d$U$ spectra taken in vortices has also been observed 
in some conventional superconductors \cite{Hess1989, Hess1990}. 
It was predicted by Caroli-de Gennes-Matricon \cite{Caroli1964} that the confinement 
of quasiparticles in vortex cores leads to confined low-energy bound states, so-called CdGM states,
with energy levels $E = \pm \mu \Delta^2/E_{\textrm{F}} \ (\mu = 1/2, 3/2 ...)$. 
Since $\Delta / E_{\textrm{F}}$ is a very small value, thermal broadening can smear out 
these discrete energy levels even at temperatures of a few Kelvin. 
Indeed, it has been shown that the X-shaped feature and the zero bias conductance peak in the vortex core 
represent the DOS that arises from many symmetric CdGM states 
when the quantum limit $T / T_{\textrm{c}}\ll\Delta/E_{\textrm{F}}$ 
is not satisfied.\cite{Klein1989, Gygi1991, Sethna1989}

%%%%%%%%%%%

\section{Conclusion}
We presented the preparation procedure of clean Nb(110) surface and studied its properties by means of STM and STS. 
We showed that in order to clean the surface from oxygen contamination, it is necessary to heat it above $T = 2400^\circ$C. 
We observed a characteristic surface state resonance at $-450$\,mV in agreement with DFT calculation, 
which is strongly suppressed over oxygen-reconstructed areas. 
A toy-model for the oxygen reconstruction on Nb(110) which is formed 
when the Nb(110) surface is heated over  $T \geqslant 2160^\circ$C was presented. 
In spatially resolved spectroscopic measurements we observed the Abrikosov flux lattice  in external magnetic fields 
and find evidence for Caroli-de Gennes-Matricon states as an enhanced peak at zero bias voltage inside a vortex core.

\begin{acknowledgments}
The work was supported by the DFG through SFB1170 ``Tocotronics'' (project C02).
\end{acknowledgments}


%merlin.mbs apsrev4-1.bst 2010-07-25 4.21a (PWD, AO, DPC) hacked
%Control: key (0)
%Control: author (72) initials jnrlst
%Control: editor formatted (1) identically to author
%Control: production of article title (-1) disabled
%Control: page (0) single
%Control: year (1) truncated
%Control: production of eprint (0) enabled
\begin{thebibliography}{0}%
\makeatletter
\providecommand \@ifxundefined [1]{%
 \@ifx{#1\undefined}
}%
\providecommand \@ifnum [1]{%
 \ifnum #1\expandafter \@firstoftwo
 \else \expandafter \@secondoftwo
 \fi
}%
\providecommand \@ifx [1]{%
 \ifx #1\expandafter \@firstoftwo
 \else \expandafter \@secondoftwo
 \fi
}%
\providecommand \natexlab [1]{#1}%
\providecommand \enquote  [1]{``#1''}%
\providecommand \bibnamefont  [1]{#1}%
\providecommand \bibfnamefont [1]{#1}%
\providecommand \citenamefont [1]{#1}%
\providecommand \href@noop [0]{\@secondoftwo}%
\providecommand \href [0]{\begingroup \@sanitize@url \@href}%
\providecommand \@href[1]{\@@startlink{#1}\@@href}%
\providecommand \@@href[1]{\endgroup#1\@@endlink}%
\providecommand \@sanitize@url [0]{\catcode `\\12\catcode `\$12\catcode
  `\&12\catcode `\#12\catcode `\^12\catcode `\_12\catcode `\%12\relax}%
\providecommand \@@startlink[1]{}%
\providecommand \@@endlink[0]{}%
\providecommand \url  [0]{\begingroup\@sanitize@url \@url }%
\providecommand \@url [1]{\endgroup\@href {#1}{\urlprefix }}%
\providecommand \urlprefix  [0]{URL }%
\providecommand \Eprint [0]{\href }%
\providecommand \doibase [0]{http://dx.doi.org/}%
\providecommand \selectlanguage [0]{\@gobble}%
\providecommand \bibinfo  [0]{\@secondoftwo}%
\providecommand \bibfield  [0]{\@secondoftwo}%
\providecommand \translation [1]{[#1]}%
\providecommand \BibitemOpen [0]{}%
\providecommand \bibitemStop [0]{}%
\providecommand \bibitemNoStop [0]{.\EOS\space}%
\providecommand \EOS [0]{\spacefactor3000\relax}%
\providecommand \BibitemShut  [1]{\csname bibitem#1\endcsname}%
\let\auto@bib@innerbib\@empty
%</preamble>
\end{thebibliography}%


\begin{thebibliography}{39}%
\makeatletter
\providecommand \@ifxundefined [1]{%
 \@ifx{#1\undefined}
}%
\providecommand \@ifnum [1]{%
 \ifnum #1\expandafter \@firstoftwo
 \else \expandafter \@secondoftwo
 \fi
}%
\providecommand \@ifx [1]{%
 \ifx #1\expandafter \@firstoftwo
 \else \expandafter \@secondoftwo
 \fi
}%
\providecommand \natexlab [1]{#1}%
\providecommand \enquote  [1]{``#1''}%
\providecommand \bibnamefont  [1]{#1}%
\providecommand \bibfnamefont [1]{#1}%
\providecommand \citenamefont [1]{#1}%
\providecommand \href@noop [0]{\@secondoftwo}%
\providecommand \href [0]{\begingroup \@sanitize@url \@href}%
\providecommand \@href[1]{\@@startlink{#1}\@@href}%
\providecommand \@@href[1]{\endgroup#1\@@endlink}%
\providecommand \@sanitize@url [0]{\catcode `\\12\catcode `\$12\catcode
  `\&12\catcode `\#12\catcode `\^12\catcode `\_12\catcode `\%12\relax}%
\providecommand \@@startlink[1]{}%
\providecommand \@@endlink[0]{}%
\providecommand \url  [0]{\begingroup\@sanitize@url \@url }%
\providecommand \@url [1]{\endgroup\@href {#1}{\urlprefix }}%
\providecommand \urlprefix  [0]{URL }%
\providecommand \Eprint [0]{\href }%
\providecommand \doibase [0]{http://dx.doi.org/}%
\providecommand \selectlanguage [0]{\@gobble}%
\providecommand \bibinfo  [0]{\@secondoftwo}%
\providecommand \bibfield  [0]{\@secondoftwo}%
\providecommand \translation [1]{[#1]}%
\providecommand \BibitemOpen [0]{}%
\providecommand \bibitemStop [0]{}%
\providecommand \bibitemNoStop [0]{.\EOS\space}%
\providecommand \EOS [0]{\spacefactor3000\relax}%
\providecommand \BibitemShut  [1]{\csname bibitem#1\endcsname}%
\let\auto@bib@innerbib\@empty
%</preamble>
\bibitem [{\citenamefont {Kitaev}(2003)}]{Kitaev2003}%
  \BibitemOpen
  \bibfield  {author} {\bibinfo {author} {\bibfnamefont {A.}~\bibnamefont
  {Kitaev}},\ }\href
  {http://www.sciencedirect.com/science/article/pii/S0003491602000180}
  {\bibfield  {journal} {\bibinfo  {journal} {Annals of Physics}\ }\textbf
  {\bibinfo {volume} {303}},\ \bibinfo {pages} {2 } (\bibinfo {year}
  {2003})}\BibitemShut {NoStop}%
\bibitem [{\citenamefont {Nayak}\ \emph {et~al.}(2008)\citenamefont {Nayak},
  \citenamefont {Simon}, \citenamefont {Stern}, \citenamefont {Freedman},\ and\
  \citenamefont {Das~Sarma}}]{Nayak2008}%
  \BibitemOpen
  \bibfield  {author} {\bibinfo {author} {\bibfnamefont {C.}~\bibnamefont
  {Nayak}}, \bibinfo {author} {\bibfnamefont {S.~H.}\ \bibnamefont {Simon}},
  \bibinfo {author} {\bibfnamefont {A.}~\bibnamefont {Stern}}, \bibinfo
  {author} {\bibfnamefont {M.}~\bibnamefont {Freedman}}, \ and\ \bibinfo
  {author} {\bibfnamefont {S.}~\bibnamefont {Das~Sarma}},\ }\href {\doibase
  10.1103/RevModPhys.80.1083} {\bibfield  {journal} {\bibinfo  {journal} {Rev.
  Mod. Phys.}\ }\textbf {\bibinfo {volume} {80}},\ \bibinfo {pages} {1083}
  (\bibinfo {year} {2008})}\BibitemShut {NoStop}%
\bibitem [{\citenamefont {Alicea}\ \emph {et~al.}(2011)\citenamefont {Alicea},
  \citenamefont {Oreg}, \citenamefont {Refael}, \citenamefont {von Oppen},\
  and\ \citenamefont {Fisher}}]{Alicea2010}%
  \BibitemOpen
  \bibfield  {author} {\bibinfo {author} {\bibfnamefont {J.}~\bibnamefont
  {Alicea}}, \bibinfo {author} {\bibfnamefont {Y.}~\bibnamefont {Oreg}},
  \bibinfo {author} {\bibfnamefont {G.}~\bibnamefont {Refael}}, \bibinfo
  {author} {\bibfnamefont {F.}~\bibnamefont {von Oppen}}, \ and\ \bibinfo
  {author} {\bibfnamefont {M.~P.~A.}\ \bibnamefont {Fisher}},\ }\href
  {http://dx.doi.org/10.1038/nphys1915} {\bibfield  {journal} {\bibinfo
  {journal} {Nat. Phys.}\ }\textbf {\bibinfo {volume} {7}},\ \bibinfo {pages}
  {412} (\bibinfo {year} {2011})}\BibitemShut {NoStop}%
\bibitem [{\citenamefont {Beenakker}(2013)}]{Beenakker2013}%
  \BibitemOpen
  \bibfield  {author} {\bibinfo {author} {\bibfnamefont {C.~W.~J.}\
  \bibnamefont {Beenakker}},\ }\href {\doibase
  http://dx.doi.org/10.1146/annurev-conmatphys-030212-184337} {\bibfield
  {journal} {\bibinfo  {journal} {Ann. Rev. Cond. Matt. Phys.}\ }\textbf
  {\bibinfo {volume} {4}},\ \bibinfo {pages} {113} (\bibinfo {year}
  {2013})}\BibitemShut {NoStop}%
\bibitem [{\citenamefont {Ivanov}(2001)}]{Ivanov2001}%
  \BibitemOpen
  \bibfield  {author} {\bibinfo {author} {\bibfnamefont {D.~A.}\ \bibnamefont
  {Ivanov}},\ }\href {\doibase http://dx.doi.org/10.1103/physrevlett.86.268}
  {\bibfield  {journal} {\bibinfo  {journal} {Phys. Rev. Lett.}\ }\textbf
  {\bibinfo {volume} {86}},\ \bibinfo {pages} {268} (\bibinfo {year}
  {2001})}\BibitemShut {NoStop}%
\bibitem [{\citenamefont {Nadj-Perge}\ \emph {et~al.}(2013)\citenamefont
  {Nadj-Perge}, \citenamefont {Drozdov}, \citenamefont {Bernevig},\ and\
  \citenamefont {Yazdani}}]{Nadj-Perge2013}%
  \BibitemOpen
  \bibfield  {author} {\bibinfo {author} {\bibfnamefont {S.}~\bibnamefont
  {Nadj-Perge}}, \bibinfo {author} {\bibfnamefont {I.~K.}\ \bibnamefont
  {Drozdov}}, \bibinfo {author} {\bibfnamefont {B.~A.}\ \bibnamefont
  {Bernevig}}, \ and\ \bibinfo {author} {\bibfnamefont {A.}~\bibnamefont
  {Yazdani}},\ }\href {\doibase 10.1103/PhysRevB.88.020407} {\bibfield
  {journal} {\bibinfo  {journal} {Phys. Rev. B}\ }\textbf {\bibinfo {volume}
  {88}},\ \bibinfo {pages} {020407} (\bibinfo {year} {2013})}\BibitemShut
  {NoStop}%
\bibitem [{\citenamefont {Nadj-Perge}\ \emph {et~al.}(2014)\citenamefont
  {Nadj-Perge}, \citenamefont {Drozdov}, \citenamefont {Li}, \citenamefont
  {Chen}, \citenamefont {Jeon}, \citenamefont {Seo}, \citenamefont {MacDonald},
  \citenamefont {Bernevig},\ and\ \citenamefont {Yazdani}}]{Yazdani2014}%
  \BibitemOpen
  \bibfield  {author} {\bibinfo {author} {\bibfnamefont {S.}~\bibnamefont
  {Nadj-Perge}}, \bibinfo {author} {\bibfnamefont {I.~K.}\ \bibnamefont
  {Drozdov}}, \bibinfo {author} {\bibfnamefont {J.}~\bibnamefont {Li}},
  \bibinfo {author} {\bibfnamefont {H.}~\bibnamefont {Chen}}, \bibinfo {author}
  {\bibfnamefont {S.}~\bibnamefont {Jeon}}, \bibinfo {author} {\bibfnamefont
  {J.}~\bibnamefont {Seo}}, \bibinfo {author} {\bibfnamefont {A.~H.}\
  \bibnamefont {MacDonald}}, \bibinfo {author} {\bibfnamefont {B.~A.}\
  \bibnamefont {Bernevig}}, \ and\ \bibinfo {author} {\bibfnamefont
  {A.}~\bibnamefont {Yazdani}},\ }\href {\doibase 10.1126/science.1259327}
  {\bibfield  {journal} {\bibinfo  {journal} {Science}\ }\textbf {\bibinfo
  {volume} {346}},\ \bibinfo {pages} {602} (\bibinfo {year}
  {2014})}\BibitemShut {NoStop}%
\bibitem [{\citenamefont {M{\'e}nard}\ \emph {et~al.}(2017)\citenamefont
  {M{\'e}nard}, \citenamefont {Guissart}, \citenamefont {Brun}, \citenamefont
  {Leriche}, \citenamefont {Trif}, \citenamefont {Debontridder}, \citenamefont
  {Demaille}, \citenamefont {Roditchev}, \citenamefont {Simon},\ and\
  \citenamefont {Cren}}]{Menard2017}%
  \BibitemOpen
  \bibfield  {author} {\bibinfo {author} {\bibfnamefont {G.~C.}\ \bibnamefont
  {M{\'e}nard}}, \bibinfo {author} {\bibfnamefont {S.}~\bibnamefont
  {Guissart}}, \bibinfo {author} {\bibfnamefont {C.}~\bibnamefont {Brun}},
  \bibinfo {author} {\bibfnamefont {R.~T.}\ \bibnamefont {Leriche}}, \bibinfo
  {author} {\bibfnamefont {M.}~\bibnamefont {Trif}}, \bibinfo {author}
  {\bibfnamefont {F.}~\bibnamefont {Debontridder}}, \bibinfo {author}
  {\bibfnamefont {D.}~\bibnamefont {Demaille}}, \bibinfo {author}
  {\bibfnamefont {D.}~\bibnamefont {Roditchev}}, \bibinfo {author}
  {\bibfnamefont {P.}~\bibnamefont {Simon}}, \ and\ \bibinfo {author}
  {\bibfnamefont {T.}~\bibnamefont {Cren}},\ }\href
  {https://doi.org/10.1038/s41467-017-02192-x} {\bibfield  {journal} {\bibinfo
  {journal} {Nat. Comm.}\ }\textbf {\bibinfo {volume} {8}},\ \bibinfo {pages}
  {2040} (\bibinfo {year} {2017})}\BibitemShut {NoStop}%
\bibitem [{\citenamefont {Kim}\ \emph {et~al.}(2018)\citenamefont {Kim},
  \citenamefont {Palacio-Morales}, \citenamefont {Posske}, \citenamefont
  {R{\'o}zsa}, \citenamefont {Palot{\'a}s}, \citenamefont {Szunyogh},
  \citenamefont {Thorwart},\ and\ \citenamefont {Wiesendanger}}]{Kim2018}%
  \BibitemOpen
  \bibfield  {author} {\bibinfo {author} {\bibfnamefont {H.}~\bibnamefont
  {Kim}}, \bibinfo {author} {\bibfnamefont {A.}~\bibnamefont
  {Palacio-Morales}}, \bibinfo {author} {\bibfnamefont {T.}~\bibnamefont
  {Posske}}, \bibinfo {author} {\bibfnamefont {L.}~\bibnamefont {R{\'o}zsa}},
  \bibinfo {author} {\bibfnamefont {K.}~\bibnamefont {Palot{\'a}s}}, \bibinfo
  {author} {\bibfnamefont {L.}~\bibnamefont {Szunyogh}}, \bibinfo {author}
  {\bibfnamefont {M.}~\bibnamefont {Thorwart}}, \ and\ \bibinfo {author}
  {\bibfnamefont {R.}~\bibnamefont {Wiesendanger}},\ }\href {\doibase
  10.1126/sciadv.aar5251} {\bibfield  {journal} {\bibinfo  {journal} {Science
  Advances}\ }\textbf {\bibinfo {volume} {4}},\ \bibinfo {pages} {eaar5251}
  (\bibinfo {year} {2018})}\BibitemShut {NoStop}%
\bibitem [{\citenamefont {Haas}(1966)}]{Haas1966}%
  \BibitemOpen
  \bibfield  {author} {\bibinfo {author} {\bibfnamefont {T.~W.}\ \bibnamefont
  {Haas}},\ }\href {\doibase http://dx.doi.org/10.1016/0039-6028(66)90033-1}
  {\bibfield  {journal} {\bibinfo  {journal} {Surf. Sci.}\ }\textbf {\bibinfo
  {volume} {5}},\ \bibinfo {pages} {345} (\bibinfo {year} {1966})}\BibitemShut
  {NoStop}%
\bibitem [{\citenamefont {Pantel}\ \emph {et~al.}(1977)\citenamefont {Pantel},
  \citenamefont {Bujor},\ and\ \citenamefont {Bardolle}}]{Pantel1977}%
  \BibitemOpen
  \bibfield  {author} {\bibinfo {author} {\bibfnamefont {R.}~\bibnamefont
  {Pantel}}, \bibinfo {author} {\bibfnamefont {M.}~\bibnamefont {Bujor}}, \
  and\ \bibinfo {author} {\bibfnamefont {J.}~\bibnamefont {Bardolle}},\ }\href
  {\doibase https://doi.org/10.1016/0039-6028(77)90103-0} {\bibfield  {journal}
  {\bibinfo  {journal} {Surf. Sci.}\ }\textbf {\bibinfo {volume} {62}},\
  \bibinfo {pages} {589 } (\bibinfo {year} {1977})}\BibitemShut {NoStop}%
\bibitem [{\citenamefont {Franchy}\ \emph {et~al.}(1996)\citenamefont
  {Franchy}, \citenamefont {Bartke},\ and\ \citenamefont
  {Gassmann}}]{Franchy1996}%
  \BibitemOpen
  \bibfield  {author} {\bibinfo {author} {\bibfnamefont {R.}~\bibnamefont
  {Franchy}}, \bibinfo {author} {\bibfnamefont {T.~U.}\ \bibnamefont {Bartke}},
  \ and\ \bibinfo {author} {\bibfnamefont {P.}~\bibnamefont {Gassmann}},\
  }\href {\doibase http://dx.doi.org/10.1016/0039-6028(96)00781-9} {\bibfield
  {journal} {\bibinfo  {journal} {Surf. Sci.}\ }\textbf {\bibinfo {volume}
  {366}},\ \bibinfo {pages} {60} (\bibinfo {year} {1996})}\BibitemShut
  {NoStop}%
\bibitem [{\citenamefont {S{\"u}rgers}\ \emph {et~al.}(2001)\citenamefont
  {S{\"u}rgers}, \citenamefont {Sch{\"o}ck},\ and\ \citenamefont
  {v.~L{\"o}hneysen}}]{Suergers2001}%
  \BibitemOpen
  \bibfield  {author} {\bibinfo {author} {\bibfnamefont {C.}~\bibnamefont
  {S{\"u}rgers}}, \bibinfo {author} {\bibfnamefont {M.}~\bibnamefont
  {Sch{\"o}ck}}, \ and\ \bibinfo {author} {\bibfnamefont {H.}~\bibnamefont
  {v.~L{\"o}hneysen}},\ }\href {\doibase
  https://doi.org/10.1016/S0039-6028(00)00908-0} {\bibfield  {journal}
  {\bibinfo  {journal} {Surf. Sci.}\ }\textbf {\bibinfo {volume} {471}},\
  \bibinfo {pages} {209 } (\bibinfo {year} {2001})}\BibitemShut {NoStop}%
\bibitem [{\citenamefont {Razinkin}\ and\ \citenamefont
  {Kuznetsov}(2010)}]{Razinkin2010}%
  \BibitemOpen
  \bibfield  {author} {\bibinfo {author} {\bibfnamefont {A.~S.}\ \bibnamefont
  {Razinkin}}\ and\ \bibinfo {author} {\bibfnamefont {M.~V.}\ \bibnamefont
  {Kuznetsov}},\ }\href {\doibase http://dx.doi.org/10.1134/s0031918x10120033}
  {\bibfield  {journal} {\bibinfo  {journal} {The Physics of Metals and
  Metallography}\ }\textbf {\bibinfo {volume} {110}},\ \bibinfo {pages} {531}
  (\bibinfo {year} {2010})}\BibitemShut {NoStop}%
\bibitem [{\citenamefont {Haas}\ \emph {et~al.}(1967)\citenamefont {Haas},
  \citenamefont {Jackson},\ and\ \citenamefont {Hooker}}]{Haas1967}%
  \BibitemOpen
  \bibfield  {author} {\bibinfo {author} {\bibfnamefont {T.~W.}\ \bibnamefont
  {Haas}}, \bibinfo {author} {\bibfnamefont {A.~G.}\ \bibnamefont {Jackson}}, \
  and\ \bibinfo {author} {\bibfnamefont {M.~P.}\ \bibnamefont {Hooker}},\
  }\href {https://doi.org/10.1063/1.1841173} {\bibfield  {journal} {\bibinfo
  {journal} {The Journal of Chemical Physics}\ }\textbf {\bibinfo {volume}
  {46}},\ \bibinfo {pages} {3025} (\bibinfo {year} {1967})}\BibitemShut
  {NoStop}%
\bibitem [{\citenamefont {Haas}(1968)}]{Haas1968}%
  \BibitemOpen
  \bibfield  {author} {\bibinfo {author} {\bibfnamefont {T.~W.}\ \bibnamefont
  {Haas}},\ }\href {\doibase 10.1063/1.1656077} {\bibfield  {journal} {\bibinfo
   {journal} {J. Appl. Phys.}\ }\textbf {\bibinfo {volume} {39}},\ \bibinfo
  {pages} {5854} (\bibinfo {year} {1968})}\BibitemShut {NoStop}%
\bibitem [{Sup()}]{SupplMat}%
  \BibitemOpen
  \href@noop {} {}\bibinfo {note} {See supplemental material for detailed
  information regarding temperature measurement, details on hydrogen
  contamination of Nb(110) surface, and a comparison of the total density of
  states in the bulk and above the surface.}\BibitemShut {Stop}%
\bibitem [{\citenamefont {Kresse}\ and\ \citenamefont
  {Furthm\"uller}(1996)}]{vasp1}%
  \BibitemOpen
  \bibfield  {author} {\bibinfo {author} {\bibfnamefont {G.}~\bibnamefont
  {Kresse}}\ and\ \bibinfo {author} {\bibfnamefont {J.}~\bibnamefont
  {Furthm\"uller}},\ }\href {\doibase 10.1103/PhysRevB.54.11169} {\bibfield
  {journal} {\bibinfo  {journal} {Phys. Rev. B}\ }\textbf {\bibinfo {volume}
  {54}},\ \bibinfo {pages} {11169} (\bibinfo {year} {1996})}\BibitemShut
  {NoStop}%
\bibitem [{vas()}]{vasp2}%
  \BibitemOpen
  \href@noop {} {}\bibinfo {howpublished} {See \url{www.vasp.at}}\BibitemShut
  {NoStop}%
\bibitem [{\citenamefont {Bl\"ochl}(1994)}]{blo}%
  \BibitemOpen
  \bibfield  {author} {\bibinfo {author} {\bibfnamefont {P.~E.}\ \bibnamefont
  {Bl\"ochl}},\ }\href {\doibase 10.1103/PhysRevB.50.17953} {\bibfield
  {journal} {\bibinfo  {journal} {Phys. Rev. B}\ }\textbf {\bibinfo {volume}
  {50}},\ \bibinfo {pages} {17953} (\bibinfo {year} {1994})}\BibitemShut
  {NoStop}%
\bibitem [{\citenamefont {Kresse}\ and\ \citenamefont {Joubert}(1999)}]{blo1}%
  \BibitemOpen
  \bibfield  {author} {\bibinfo {author} {\bibfnamefont {G.}~\bibnamefont
  {Kresse}}\ and\ \bibinfo {author} {\bibfnamefont {D.}~\bibnamefont
  {Joubert}},\ }\href {\doibase 10.1103/PhysRevB.59.1758} {\bibfield  {journal}
  {\bibinfo  {journal} {Phys. Rev. B}\ }\textbf {\bibinfo {volume} {59}},\
  \bibinfo {pages} {1758} (\bibinfo {year} {1999})}\BibitemShut {NoStop}%
\bibitem [{\citenamefont {Perdew}\ \emph {et~al.}(1996)\citenamefont {Perdew},
  \citenamefont {Burke},\ and\ \citenamefont {Ernzerhof}}]{PBE}%
  \BibitemOpen
  \bibfield  {author} {\bibinfo {author} {\bibfnamefont {J.~P.}\ \bibnamefont
  {Perdew}}, \bibinfo {author} {\bibfnamefont {K.}~\bibnamefont {Burke}}, \
  and\ \bibinfo {author} {\bibfnamefont {M.}~\bibnamefont {Ernzerhof}},\ }\href
  {\doibase 10.1103/PhysRevLett.77.3865} {\bibfield  {journal} {\bibinfo
  {journal} {Phys. Rev. Lett.}\ }\textbf {\bibinfo {volume} {77}},\ \bibinfo
  {pages} {3865} (\bibinfo {year} {1996})}\BibitemShut {NoStop}%
\bibitem [{\citenamefont {Perdew}\ \emph {et~al.}(1997)\citenamefont {Perdew},
  \citenamefont {Burke},\ and\ \citenamefont {Ernzerhof}}]{PBEerr}%
  \BibitemOpen
  \bibfield  {author} {\bibinfo {author} {\bibfnamefont {J.~P.}\ \bibnamefont
  {Perdew}}, \bibinfo {author} {\bibfnamefont {K.}~\bibnamefont {Burke}}, \
  and\ \bibinfo {author} {\bibfnamefont {M.}~\bibnamefont {Ernzerhof}},\ }\href
  {\doibase 10.1103/PhysRevLett.78.1396} {\bibfield  {journal} {\bibinfo
  {journal} {Phys. Rev. Lett.}\ }\textbf {\bibinfo {volume} {78}},\ \bibinfo
  {pages} {1396} (\bibinfo {year} {1997})}\BibitemShut {NoStop}%
\bibitem [{\citenamefont {Haas}\ \emph {et~al.}(2009)\citenamefont {Haas},
  \citenamefont {Tran},\ and\ \citenamefont {Blaha}}]{Haas2009}%
  \BibitemOpen
  \bibfield  {author} {\bibinfo {author} {\bibfnamefont {P.}~\bibnamefont
  {Haas}}, \bibinfo {author} {\bibfnamefont {F.}~\bibnamefont {Tran}}, \ and\
  \bibinfo {author} {\bibfnamefont {P.}~\bibnamefont {Blaha}},\ }\href
  {\doibase 10.1103/PhysRevB.79.085104} {\bibfield  {journal} {\bibinfo
  {journal} {Phys. Rev. B}\ }\textbf {\bibinfo {volume} {79}},\ \bibinfo
  {pages} {085104} (\bibinfo {year} {2009})}\BibitemShut {NoStop}%
\bibitem [{\citenamefont {Kr\"oger}\ \emph {et~al.}(2000)\citenamefont
  {Kr\"oger}, \citenamefont {Greber},\ and\ \citenamefont
  {Osterwalder}}]{KGO2000}%
  \BibitemOpen
  \bibfield  {author} {\bibinfo {author} {\bibfnamefont {J.}~\bibnamefont
  {Kr\"oger}}, \bibinfo {author} {\bibfnamefont {T.}~\bibnamefont {Greber}}, \
  and\ \bibinfo {author} {\bibfnamefont {J.}~\bibnamefont {Osterwalder}},\
  }\href {\doibase 10.1103/PhysRevB.61.14146} {\bibfield  {journal} {\bibinfo
  {journal} {Phys. Rev. B}\ }\textbf {\bibinfo {volume} {61}},\ \bibinfo
  {pages} {14146} (\bibinfo {year} {2000})}\BibitemShut {NoStop}%
\bibitem [{\citenamefont {Bode}\ \emph {et~al.}(2007)\citenamefont {Bode},
  \citenamefont {Krause}, \citenamefont {Berbil-Bautista}, \citenamefont
  {Heinze},\ and\ \citenamefont {Wiesendanger}}]{BKB2007}%
  \BibitemOpen
  \bibfield  {author} {\bibinfo {author} {\bibfnamefont {M.}~\bibnamefont
  {Bode}}, \bibinfo {author} {\bibfnamefont {S.}~\bibnamefont {Krause}},
  \bibinfo {author} {\bibfnamefont {L.}~\bibnamefont {Berbil-Bautista}},
  \bibinfo {author} {\bibfnamefont {S.}~\bibnamefont {Heinze}}, \ and\ \bibinfo
  {author} {\bibfnamefont {R.}~\bibnamefont {Wiesendanger}},\ }\href {\doibase
  https://doi.org/10.1016/j.susc.2007.06.017} {\bibfield  {journal} {\bibinfo
  {journal} {Surf. Sci.}\ }\textbf {\bibinfo {volume} {601}},\ \bibinfo {pages}
  {3308 } (\bibinfo {year} {2007})}\BibitemShut {NoStop}%
\bibitem [{\citenamefont {Zakeri}\ \emph {et~al.}(2010)\citenamefont {Zakeri},
  \citenamefont {Peixoto}, \citenamefont {Zhang}, \citenamefont {Prokop},\ and\
  \citenamefont {Kirschner}}]{ZPZ2010}%
  \BibitemOpen
  \bibfield  {author} {\bibinfo {author} {\bibfnamefont {K.}~\bibnamefont
  {Zakeri}}, \bibinfo {author} {\bibfnamefont {T.}~\bibnamefont {Peixoto}},
  \bibinfo {author} {\bibfnamefont {Y.}~\bibnamefont {Zhang}}, \bibinfo
  {author} {\bibfnamefont {J.}~\bibnamefont {Prokop}}, \ and\ \bibinfo {author}
  {\bibfnamefont {J.}~\bibnamefont {Kirschner}},\ }\href {\doibase
  https://doi.org/10.1016/j.susc.2009.10.020} {\bibfield  {journal} {\bibinfo
  {journal} {Surf. Sci.}\ }\textbf {\bibinfo {volume} {604}},\ \bibinfo {pages}
  {L1 } (\bibinfo {year} {2010})}\BibitemShut {NoStop}%
\bibitem [{\citenamefont {Tafen}\ and\ \citenamefont {Gao}(2013)}]{Tafen2013}%
  \BibitemOpen
  \bibfield  {author} {\bibinfo {author} {\bibfnamefont {D.~N.}\ \bibnamefont
  {Tafen}}\ and\ \bibinfo {author} {\bibfnamefont {M.~C.}\ \bibnamefont
  {Gao}},\ }\href {\doibase http://dx.doi.org/10.1007/s11837-013-0735-8}
  {\bibfield  {journal} {\bibinfo  {journal} {{JOM}}\ }\textbf {\bibinfo
  {volume} {65}},\ \bibinfo {pages} {1473} (\bibinfo {year}
  {2013})}\BibitemShut {NoStop}%
\bibitem [{\citenamefont {Ukraintsev}(1996)}]{Ukr1996}%
  \BibitemOpen
  \bibfield  {author} {\bibinfo {author} {\bibfnamefont {V.~A.}\ \bibnamefont
  {Ukraintsev}},\ }\href {\doibase 10.1103/PhysRevB.53.11176} {\bibfield
  {journal} {\bibinfo  {journal} {Phys. Rev. B}\ }\textbf {\bibinfo {volume}
  {53}},\ \bibinfo {pages} {11176} (\bibinfo {year} {1996})}\BibitemShut
  {NoStop}%
\bibitem [{\citenamefont {Smith}(1981)}]{Smith1981}%
  \BibitemOpen
  \bibfield  {author} {\bibinfo {author} {\bibfnamefont {R.~J.}\ \bibnamefont
  {Smith}},\ }\href {\doibase http://dx.doi.org/10.1016/0038-1098(81)91086-3}
  {\bibfield  {journal} {\bibinfo  {journal} {Sol. State Comm.}\ }\textbf
  {\bibinfo {volume} {37}},\ \bibinfo {pages} {725} (\bibinfo {year}
  {1981})}\BibitemShut {NoStop}%
\bibitem [{\citenamefont {Dynes}\ \emph {et~al.}(1978)\citenamefont {Dynes},
  \citenamefont {Narayanamurti},\ and\ \citenamefont {Garno}}]{Dynes1978}%
  \BibitemOpen
  \bibfield  {author} {\bibinfo {author} {\bibfnamefont {R.~C.}\ \bibnamefont
  {Dynes}}, \bibinfo {author} {\bibfnamefont {V.}~\bibnamefont
  {Narayanamurti}}, \ and\ \bibinfo {author} {\bibfnamefont {J.~P.}\
  \bibnamefont {Garno}},\ }\href {\doibase
  http://dx.doi.org/10.1103/physrevlett.41.1509} {\bibfield  {journal}
  {\bibinfo  {journal} {Phys. Rev. Lett.}\ }\textbf {\bibinfo {volume} {41}},\
  \bibinfo {pages} {1509} (\bibinfo {year} {1978})}\BibitemShut {NoStop}%
\bibitem [{\citenamefont {Finnemore}\ \emph {et~al.}(1966)\citenamefont
  {Finnemore}, \citenamefont {Stromberg},\ and\ \citenamefont
  {Swenson}}]{Finnemore1966}%
  \BibitemOpen
  \bibfield  {author} {\bibinfo {author} {\bibfnamefont {D.~K.}\ \bibnamefont
  {Finnemore}}, \bibinfo {author} {\bibfnamefont {T.~F.}\ \bibnamefont
  {Stromberg}}, \ and\ \bibinfo {author} {\bibfnamefont {C.~A.}\ \bibnamefont
  {Swenson}},\ }\href {\doibase http://dx.doi.org/10.1103/physrev.149.231}
  {\bibfield  {journal} {\bibinfo  {journal} {Phys. Rev.}\ }\textbf {\bibinfo
  {volume} {149}},\ \bibinfo {pages} {231} (\bibinfo {year}
  {1966})}\BibitemShut {NoStop}%
\bibitem [{\citenamefont {Karasik}\ and\ \citenamefont
  {Shebalin}(1970)}]{Karasik1970}%
  \BibitemOpen
  \bibfield  {author} {\bibinfo {author} {\bibfnamefont {V.~R.}\ \bibnamefont
  {Karasik}}\ and\ \bibinfo {author} {\bibfnamefont {I.~Y.}\ \bibnamefont
  {Shebalin}},\ }\href
  {http://www.jetp.ac.ru/cgi-bin/e/index/e/30/6/p1068?a=list} {\bibfield
  {journal} {\bibinfo  {journal} {Sov. Phys. JETP}\ }\textbf {\bibinfo {volume}
  {30}},\ \bibinfo {pages} {1068} (\bibinfo {year} {1970})}\BibitemShut
  {NoStop}%
\bibitem [{\citenamefont {Hess}\ \emph {et~al.}(1989)\citenamefont {Hess},
  \citenamefont {Robinson}, \citenamefont {Dynes}, \citenamefont {Valles},\
  and\ \citenamefont {Waszczak}}]{Hess1989}%
  \BibitemOpen
  \bibfield  {author} {\bibinfo {author} {\bibfnamefont {H.~F.}\ \bibnamefont
  {Hess}}, \bibinfo {author} {\bibfnamefont {R.~B.}\ \bibnamefont {Robinson}},
  \bibinfo {author} {\bibfnamefont {R.~C.}\ \bibnamefont {Dynes}}, \bibinfo
  {author} {\bibfnamefont {J.~M.}\ \bibnamefont {Valles}}, \ and\ \bibinfo
  {author} {\bibfnamefont {J.~V.}\ \bibnamefont {Waszczak}},\ }\href {\doibase
  10.1103/PhysRevLett.62.214} {\bibfield  {journal} {\bibinfo  {journal} {Phys.
  Rev. Lett.}\ }\textbf {\bibinfo {volume} {62}},\ \bibinfo {pages} {214}
  (\bibinfo {year} {1989})}\BibitemShut {NoStop}%
\bibitem [{\citenamefont {Hess}\ \emph {et~al.}(1990)\citenamefont {Hess},
  \citenamefont {Robinson},\ and\ \citenamefont {Waszczak}}]{Hess1990}%
  \BibitemOpen
  \bibfield  {author} {\bibinfo {author} {\bibfnamefont {H.~F.}\ \bibnamefont
  {Hess}}, \bibinfo {author} {\bibfnamefont {R.~B.}\ \bibnamefont {Robinson}},
  \ and\ \bibinfo {author} {\bibfnamefont {J.~V.}\ \bibnamefont {Waszczak}},\
  }\href {\doibase http://dx.doi.org/10.1103/physrevlett.64.2711} {\bibfield
  {journal} {\bibinfo  {journal} {Phys. Rev. Lett.}\ }\textbf {\bibinfo
  {volume} {64}},\ \bibinfo {pages} {2711} (\bibinfo {year}
  {1990})}\BibitemShut {NoStop}%
\bibitem [{\citenamefont {Caroli}\ \emph {et~al.}(1964)\citenamefont {Caroli},
  \citenamefont {Gennes},\ and\ \citenamefont {Matricon}}]{Caroli1964}%
  \BibitemOpen
  \bibfield  {author} {\bibinfo {author} {\bibfnamefont {C.}~\bibnamefont
  {Caroli}}, \bibinfo {author} {\bibfnamefont {P.~G.~D.}\ \bibnamefont
  {Gennes}}, \ and\ \bibinfo {author} {\bibfnamefont {J.}~\bibnamefont
  {Matricon}},\ }\href {\doibase
  http://dx.doi.org/10.1016/0031-9163(64)90375-0} {\bibfield  {journal}
  {\bibinfo  {journal} {Phys. Lett.}\ }\textbf {\bibinfo {volume} {9}},\
  \bibinfo {pages} {307} (\bibinfo {year} {1964})}\BibitemShut {NoStop}%
\bibitem [{\citenamefont {Klein}(1989)}]{Klein1989}%
  \BibitemOpen
  \bibfield  {author} {\bibinfo {author} {\bibfnamefont {U.}~\bibnamefont
  {Klein}},\ }\href {\doibase http://dx.doi.org/10.1103/physrevb.40.6601}
  {\bibfield  {journal} {\bibinfo  {journal} {Phys. Rev. B}\ }\textbf {\bibinfo
  {volume} {40}},\ \bibinfo {pages} {6601} (\bibinfo {year}
  {1989})}\BibitemShut {NoStop}%
\bibitem [{\citenamefont {Gygi}\ and\ \citenamefont
  {Schl{\"u}ter}(1991)}]{Gygi1991}%
  \BibitemOpen
  \bibfield  {author} {\bibinfo {author} {\bibfnamefont {F.}~\bibnamefont
  {Gygi}}\ and\ \bibinfo {author} {\bibfnamefont {M.}~\bibnamefont
  {Schl{\"u}ter}},\ }\href {\doibase 10.1103/physrevb.43.7609} {\bibfield
  {journal} {\bibinfo  {journal} {Phys. Rev. B}\ }\textbf {\bibinfo {volume}
  {43}},\ \bibinfo {pages} {7609} (\bibinfo {year} {1991})}\BibitemShut
  {NoStop}%
\bibitem [{\citenamefont {Shore}\ \emph {et~al.}(1989)\citenamefont {Shore},
  \citenamefont {Huang}, \citenamefont {Dorsey},\ and\ \citenamefont
  {Sethna}}]{Sethna1989}%
  \BibitemOpen
  \bibfield  {author} {\bibinfo {author} {\bibfnamefont {J.~D.}\ \bibnamefont
  {Shore}}, \bibinfo {author} {\bibfnamefont {M.}~\bibnamefont {Huang}},
  \bibinfo {author} {\bibfnamefont {A.~T.}\ \bibnamefont {Dorsey}}, \ and\
  \bibinfo {author} {\bibfnamefont {J.~P.}\ \bibnamefont {Sethna}},\ }\href
  {\doibase http://dx.doi.org/10.1103/physrevlett.62.3089} {\bibfield
  {journal} {\bibinfo  {journal} {Phys. Rev. Lett.}\ }\textbf {\bibinfo
  {volume} {62}},\ \bibinfo {pages} {3089} (\bibinfo {year}
  {1989})}\BibitemShut {NoStop}%
\end{thebibliography}
\end{document}